\documentstyle[twocolumn,aps,psfig,epsf]{revtex}

\begin{document}
\title{Segregation in hard sphere mixtures under
gravity. An extension of Edwards approach with two thermodynamical
parameters}
\author{Mario Nicodemi$^{a,b}$, Annalisa Fierro$^a$ and Antonio Coniglio$^a$}
\address{$^{a)}$
Dipartimento di Fisica, Universit\'{a} ``Federico II'', INFM and INFN
Napoli, Via Cintia, 80126 Napoli, Italy\\ $^{b)}$ Department of
Mathematics, Imperial College, London, SW7 2BZ, U.K.\\ }
\maketitle

\begin{abstract}
We study segregation patterns in a hard sphere binary model under gravity
subject to sequences of taps. We discuss the appearance of the ``Brazil
nut'' effect (where large particles move up) and the ``reverse
Brazil nut'' effects in the stationary states reached by ``tap''
dynamics. In particular, we show that the stationary state depends 
only on two thermodynamical quantities: the gravitational energy of 
the first and of the second species and not on the sample history. 
To describe the properties of the system, we generalize Edwards' approach 
by introducing a canonical distribution characterized by two 
configurational temperatures, conjugate to the energies of the two species. 
This is supported by Monte Carlo 
calculations showing that the average of several quantities over the tap
dynamics and over such distribution coincide. The segregation problem can
then be understood as an equilibrium statistical mechanics problem 
with two control parameters. 
\end{abstract}

\pagestyle{myheadings}

\vskip1pc 

Segregation is an ubiquitous and intriguing phenomenon observed in
vibrated granular mixtures such as powders or sand: in presence of shaking
the system is not randomized, but its components tend to separate
\cite{rev_segr,makse1}. 
An example is the so called ``Brazil nut'' effect
(BNE) where, under shaking, large particles rise to the top and small
particles move to the bottom of the container. Several mechanisms have
been proposed to explain these phenomena, as for instance ``percolation''
\cite{rosato}, geometrical reorganization \cite{bridgewater,duran},
convection \cite{knight}, ``condensation'' \cite{luding} or
inertia \cite{shinbrot}. Interestingly, by changing grains sizes or mass ratio 
a crossover towards a ``reverse Brazil nut'' effect (RBNE) was more recently 
discovered \cite{luding} where small particles segregates to the top and
large particles to the bottom.
On the whole the criteria to predict
segregation in a mixture, an issue of large practical relevance, are still
largely unknown \cite{rev_segr}.

In the perspective of the Statistical Mechanics approach originally
proposed by Edwards to describe powders \cite{Edwards1},
we study here segregation patterns,
in absence of convection mechanisms, of a binary hard sphere model
under gravity subject to sequences of ``taps''.
We show that a Statistical Mechanics description of segregation is
indeed possible by the introduction of two ``configurational temperatures''
which characterize the macroscopic status of the system.

In Edwards' approach granular media at rest (i.e., when grains have no 
kinetic energy and, thus, the bath temperature is zero) can be described in
terms of Statistical Mechanics where a single control parameter, called
compactivity, plays the role of the temperature of usual thermal systems
\cite{Edwards1,Dean,fnc,brey,kurchan}.
We show that the stationary states of the present mixture are indeed
independent on the sample history as in a ``thermodynamics'' system,
but are characterized by two control parameters, such as
the gravitational energy of the first and of the second species.
We postulate that the probability distribution of the system in the
stationary state can be obtained by the principle of maximum entropy under
the constraint that the two energies are independently fixed.
This leads to a Gibbs canonical
distribution characterized by two configurational temperatures, conjugate
to the energies of the two species, generalizing in this way Edwards
approach \cite{lefevre} . The postulate is supported by Monte Carlo
simulations showing for several quantities that statistical averages
over the tap dynamics coincide with those over such an ensemble distribution.
The segregation problem can then be studied as standard Statistical Mechanics
problems using such a distribution and, for instance,
a segregation ``phase'' diagram can be theoretically derived.

\begin{figure}[ht]
\centerline{
\hspace{-2.5cm}
\psfig{figure=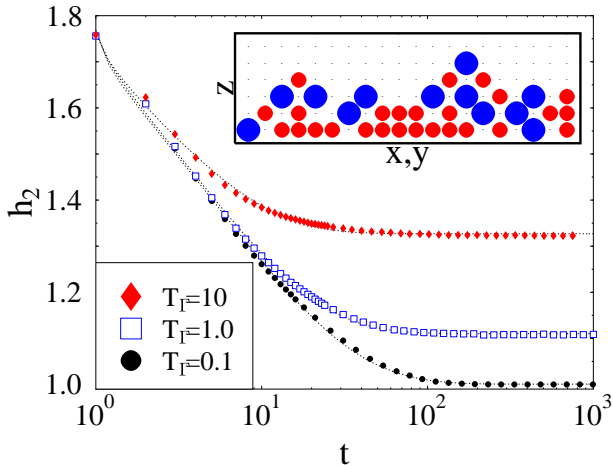,height=9.5cm,angle=-90}}
\vspace{-1.5cm}
\caption{{\bf Main frame} The relaxation of large grains average height,
$h_2$, as a function of the number of taps, $t$, in taps sequences
with the shown shaking amplitudes, $T_{\Gamma}$
(tap duration, $\protect\tau_0=10$).
The superimposed curves are stretched exponential fits.
{\bf Inset} A schematic picture of
the 3D lattice hard sphere mixture considered here.}
\label{fig_h2_t}
\end{figure}

We consider a binary hard sphere system made of two species 1 (small) and
2 (large) with grain diameters $a_0$ and $\sqrt{2} a_0$, under gravity
on a cubic lattice of spacing $a_0=1$ (see inset of Fig.\ref{fig_h2_t}).
The hard core
potential ${\cal H}_{hc}$ is such that two nearest neighbor particles
cannot overlap. This implies that only pairs of small particles can be
nearest neighbors on the lattice. The system overall Hamiltonian is:
\begin{equation}
{\cal H}={\cal H}_{hc}+ m_1gH_1 + m_2gH_2,
\label{H}
\end{equation}
where $H_1=\sum_{i}^{(1)}z_{i}$ and  $H_2=\sum_{i}^{(2)}z_{i}$, the
height of site $i$ is $z_i$ and the two sums are over all particles of
species 1 and 2 respectively. We set the units such that the two kinds 
of grain have masses $m_1=1$ and $m_2=2$, and the gravity acceleration 
is $g=1$. In the above units, the gravitational
energies in a given configuration are thus $E_1=H_1$ and $E_2=2H_2$.

Grains are confined in a box of linear size $L$ between hard walls and
periodic boundary conditions in the horizontal directions. The system is
subject to a Monte Carlo dynamics made of sequences of ``taps''
\cite{NCH,Dean,fnc}.
During each ``tap'' the system evolves for a time $\tau_0$ (the
``tap duration'', in units of lattice sweeps) in presence of a finite
thermal bath of temperature $T_{\Gamma}$; afterwards it is suddenly frozen
at zero temperature in one of its stable states. These states correspond
to a local minimum or saddle of the energy, such that any particle movement
does not decrease the energy, and have been called ``inherent states''
\cite{fnc} 
in analogy with the terminology of glasses \cite{Stillinger}. After
each tap, when the system is at rest, we record the quantity of interest,
which are a function of the number of ``taps'', $t$.

\begin{figure}[ht]
\centerline{
\hspace{-2.5cm}
\psfig{figure=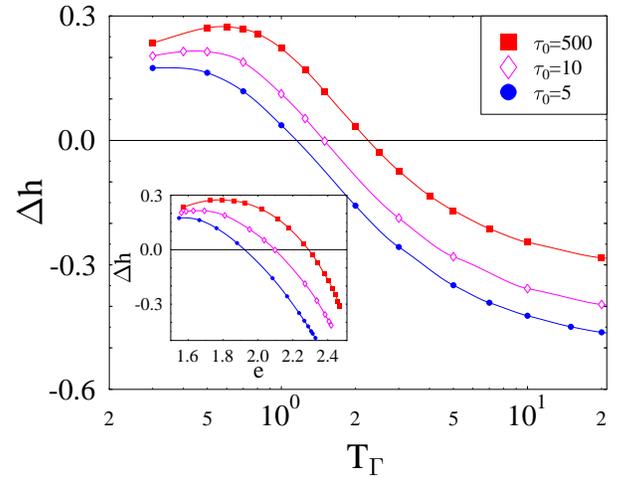,height=9.5cm,angle=-90}}
\vspace{-1.5cm}
\caption{{\bf Main frame} The difference of the average heights of small
and large grains, $\Delta h=h_1-h_2$, measured at stationarity is
plotted as a function of tap amplitude, $T_{\Gamma}$. The three sets of
points correspond to the shown tap durations, $\protect\tau_0$.
At high $T_{\Gamma}$ larger grains are found above the smaller, i.e,
$\Delta h<0$, as in the Brazil nut effect (BNE).
Below a  $T^*_{\Gamma}(\tau_0)$ the opposite is found
(Reverse Brazil nut effect, RBNE).
{\bf Inset}
The $\Delta h$ data of the main frame are plotted as a function of the
corresponding average energy, $e$. The three sets of data do not collapse on
a single master function, showing that a single macroscopic observable, such
as $e$, doesn't characterize unambiguously the system status.
}
\label{fig_dh}
\end{figure}

$N_1$ grains of type 1 and $N_2$ grains of type 2 are initially prepared
in a random loose stable pack \cite{nota_num_p}. Under ``shaking'' the
system is let to approach a stationary state for each value of the 
tap parameters $T_{\Gamma}$ and $\tau_0$, as shown in Fig.\ref{fig_h2_t}
for the large grains height $h_2(t)= H_2(t)/N_2$.
In Fig.\ref{fig_dh}, we plot as function of $ T_{\Gamma}$ (for several values
of $\tau_0$) the asymptotic value of the
{\em vertical} segregation parameter, i.e.,
the difference of the average heights of the small and large grains at
stationarity, $\Delta h(T_{\Gamma},\tau_0)\equiv h_1-h_2$.
Here $h_1$ and $h_2$ are the average of $H_1/N_1$ and $H_2/N_2$
over the tap dynamics in the stationary state.

Fig.\ref{fig_dh} shows that at high $T_{\Gamma}$ the
Brazil Nut Effect (BNE) of granular media is observed:
``large" grains are found on the top of the system (i.e., $\Delta h <0 $).
For small values of $T_{\Gamma}$, below a crossover amplitude,
$T^*_{\Gamma}(\tau_0)$, a smooth crossover to the Reverse
Brazil Nut Effect (RBNE) is instead recorded, with $\Delta h >0$. 
The RBNE was first discovered in recent Molecular Dynamics 
simulations \cite{luding}.
Our aim below is to show that the properties of these segregation patterns
can be explained in terms of a statistical mechanics approach to the problem.

\bigskip

The results given in the main panel of Fig.~\ref{fig_dh}
apparently show that $T_{\Gamma}$ is not a
right ``thermodynamical" parameter, since sequences of ``taps" with
different $\tau_0$ give different values for the system observables.
However, if the stationary states corresponding to different tap dynamics
(i.e., different $T_{\Gamma}$ and $\tau_0$), are indeed characterized by a
single ``thermodynamical" parameter \'{a} la Edwards, the curves of
Fig.\ref{fig_dh} should collapse on an universal master function when
$\Delta h(T_{\Gamma},\tau_0)$ is parametrically plotted as function of an
other macroscopic observable such as the average energy,
$e(T_{\Gamma},\tau_0)=(E_1 + E_2)/N={N_1\over N}h_1+{2N_2\over N}h_2$ 
(where $N=N_1+N_2$). 
This collapse of data, actually found in other situations \cite{Dean,fnc}, 
is clearly not observed here, as it is apparent in the inset of
Fig.\ref{fig_dh} (see also \cite{berg}).

\begin{figure}[ht]
\centerline{
\hspace{-2.5cm}
\psfig{figure=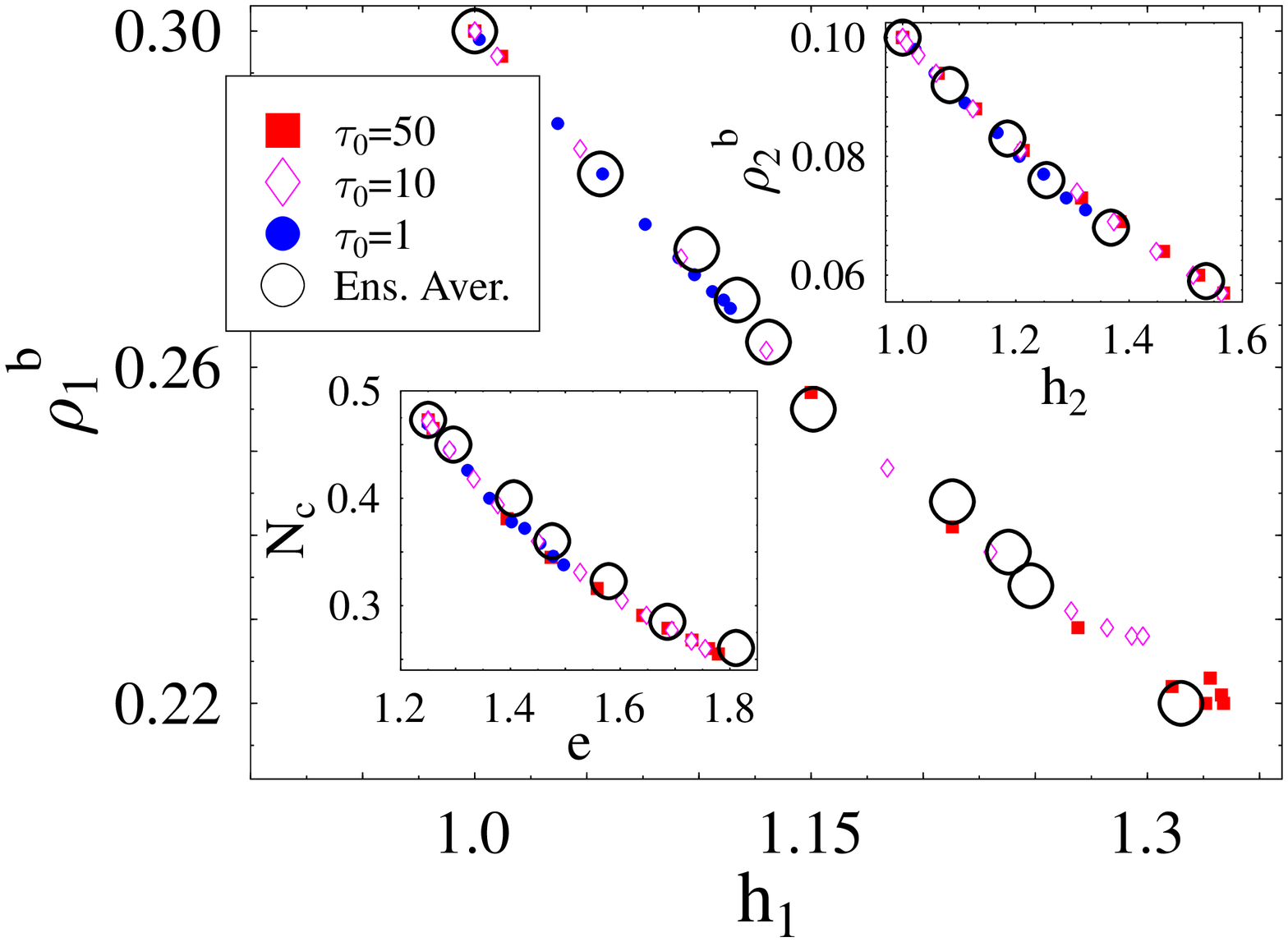,height=9.5cm,angle=-90}}
\vspace{-1.5cm}
\caption{{\bf Main frame} The average density of small grains on the
box bottom layer, $\protect\rho_1^{b}$, measured at
stationarity as a function of the height of small particles, $h_1$. Data
corresponding to different $T_{\Gamma}$ and $\protect\tau_0$ approximately
scale on a single master function.
The empty circles are the corresponding values obtained by ensemble
average with the two temperatures Gibbs measure proposed in the text.
{\bf Upper inset} The average density of
large grains on the box bottom layer, $\protect\rho_2^{b}$, obtained for
different $T_{\Gamma}$ and $\protect\tau_0$, scale almost on a single master
function when plotted as a function of the large grains height, $h_2$.
{\bf Lower inset} The average number of contacts between large grains
per particle, $N_c$, obtained for different $T_{\Gamma}$ and $\protect\tau_0$,
scale on a single master function when plotted as a function of the system
energy, $e$. }
\label{fig_qr2nc}
\end{figure}

We show, instead, that two macroscopic quantities, such as $h_1$ and $h_2$, 
may be sufficient to characterize uniquely the stationary state of the system. 
Namely, we show that any macroscopic
quantity $Q$, averaged over the tap dynamics in
the stationary state, is only dependent on $h_1$ and $h_2$,
i.e., $Q= Q(h_1,h_2)$. We have checked that this is the case for
several independent observables, such as
the number of contacts between large particles, $N_{c}$, the density
of  small and large particles on the bottom layer, $\rho_{1}^{b}$ and
$\rho_{2}^{b}$, and others.
As shown in Fig.\ref {fig_qr2nc}, we find, in particular,
with good approximation that: $N_{c}\simeq N_{c}(e)=N_{c}({N_1\over N}h_1+
{2N_2\over N}h_2)$,
$\rho_{2}^{b}\simeq\rho_{2}^{b}(h_{2})$, $
\rho _{1}^{b}\simeq \rho _{1}^{b}(h_{1})$.
Therefore we need both $h_{1}$ and $h_{2}$ (or equivalently $E_1$ and $E_2$) 
to characterize unambiguously the state of the system 
independently on the previous history (i.e., in our case independently on the
particular tapping parameters $T_{\Gamma}$ and $\tau_0$).

In this respect the stationary state can be genuinely considered as a
``thermodynamical state", and consequently one can attempt to construct an
equilibrium statistical mechanics, as originally suggested by Edwards.
This will allow to
calculate any quantity in the stationary state using an appropriate
ensemble, irrespective of the particular history.

In the present paper we generalize Edwards' approach by 
introducing a canonical distribution characterized by two configurational 
temperatures, conjugate to the energies of the two species, and verify by  
Monte Carlo calculations that the average of several quantities over the tap 
dynamics and over such distribution coincide. 
More precisely we have to find the probability distribution, $P_r$, of
observing the system in the generic inherent state $r$ corresponding to an
energy $E_{1r}$ for the small particles and $E_{2r}$ for the large
particles (see \cite{fnc}).
To answer this, we assume that the microscopic distribution is given by the
principle of maximum entropy $S=-\sum_r P_r\ln P_r$ under the condition
that the average energy $E_1 =\sum_r P_r E_{1r} $ and
$E_2 =\sum_r P_r E_{2r} $ are independently fixed.
This can be done by introducing two Lagrange multipliers
$\beta_1$ and $\beta_2$, which are  determined by the constraint on $ E_1$
and $E_2$, and can be considered as the ``inverse configurational
temperature'' of species 1 and 2 \cite{fnc}. 
This procedure leads to the Gibbs result:
\begin{eqnarray}
P_r=\frac{e^{-\beta_1 E_{1r} - \beta_2 E_{2r}}}{Z}
\label{pr}
\end{eqnarray}
where
$Z=\sum_r \exp\left[-\beta_1 E_{1r} - \beta_2 E_{2r}\right]$
and, in the thermodynamic limit, the entropy, $S$, is given by:
\begin{eqnarray}
S = \ln \Omega (E_1,E_2),
\label{omega}
\end{eqnarray}
and $\beta_1$ and $\beta_2$:
\begin{eqnarray}
\beta_1 = \frac{\partial \ln \Omega (E_1,E_2) }{\partial E_1},\quad
\beta_2 = \frac{\partial \ln \Omega (E_1,E_2) }{\partial E_2}.
\label{beta}
\end{eqnarray}
Here $\Omega (E_1,E_2)$ is the number of inherent states corresponding to
energy $E_1$ and $E_2$.

The hypothesis that the ensemble distribution at stationarity is given
by eq.(\ref{pr}) can be tested as follows. We have to check that
the time average of any quantity,
$Q(h_{1},h_{2})$, as
recorded during the taps sequences in a stationary state characterized
by given values  $h_{1}$ and $h_{2}$, must coincide with the ensemble
average, $\langle Q\rangle (h_{1},h_{2})$, over the distribution
eq.(\ref{pr}). 
To this aim, we have calculated the
ensemble averages $\langle N_{c}\rangle $,  $\langle \rho_{2}^{b}\rangle$,
$\langle \rho _{1}^{b}\rangle $, 
for different  values of $\beta_1$ and $\beta_2$; then we have expressed
parametrically  $\langle N_{c}\rangle $, $\langle
\rho _{2}^{b}\rangle $, $\langle \rho _{1}^{b}\rangle $,
as function of the average of $h_{1}$ and $h_{2}$,
and compared them with the corresponding quantities, $N_{c}$, $\rho_{1}^{b}$
and $\rho _{2}^{b}$, averaged over the tap dynamics.
The two sets of data are plotted in Fig.\ref{fig_qr2nc} showing a
rather good agreement (notice, there are no adjustable parameters).
Technically, in order to calculate the ensemble averages we simulated, 
with standard Monte Carlo methods, the model with $\cal H$ from eq.(\ref{H})
where we impose that the only accessible states are the inherent states. 
This can be done by using a configurational Hamiltonian
\begin{equation} 
{\cal H}_{conf}={\cal H}_{hc}+ \beta_1 m_1gH_1 + \beta_2 m_2gH_2- \ln \Pi, 
\label{effective}
\end{equation}
where
$\Pi=1$ or $0$ depending whether the configuration is an inherent state or not. In this way the weight $e^{-{\cal H}_{conf}}$ gives the ensamble distribution 
(\ref{pr}) (for further details see \cite{fnc,kurchan}). 
The discovery of two configurational temperatures may suggest 
the existence of more than one ``dynamical temperature'' in the 
off-equilibrium fluctuation-dissipation relation approach of 
\cite{kurchan,mimmo}. 

\begin{figure}[ht]
\centerline{
\hspace{-2.5cm}
\psfig{figure=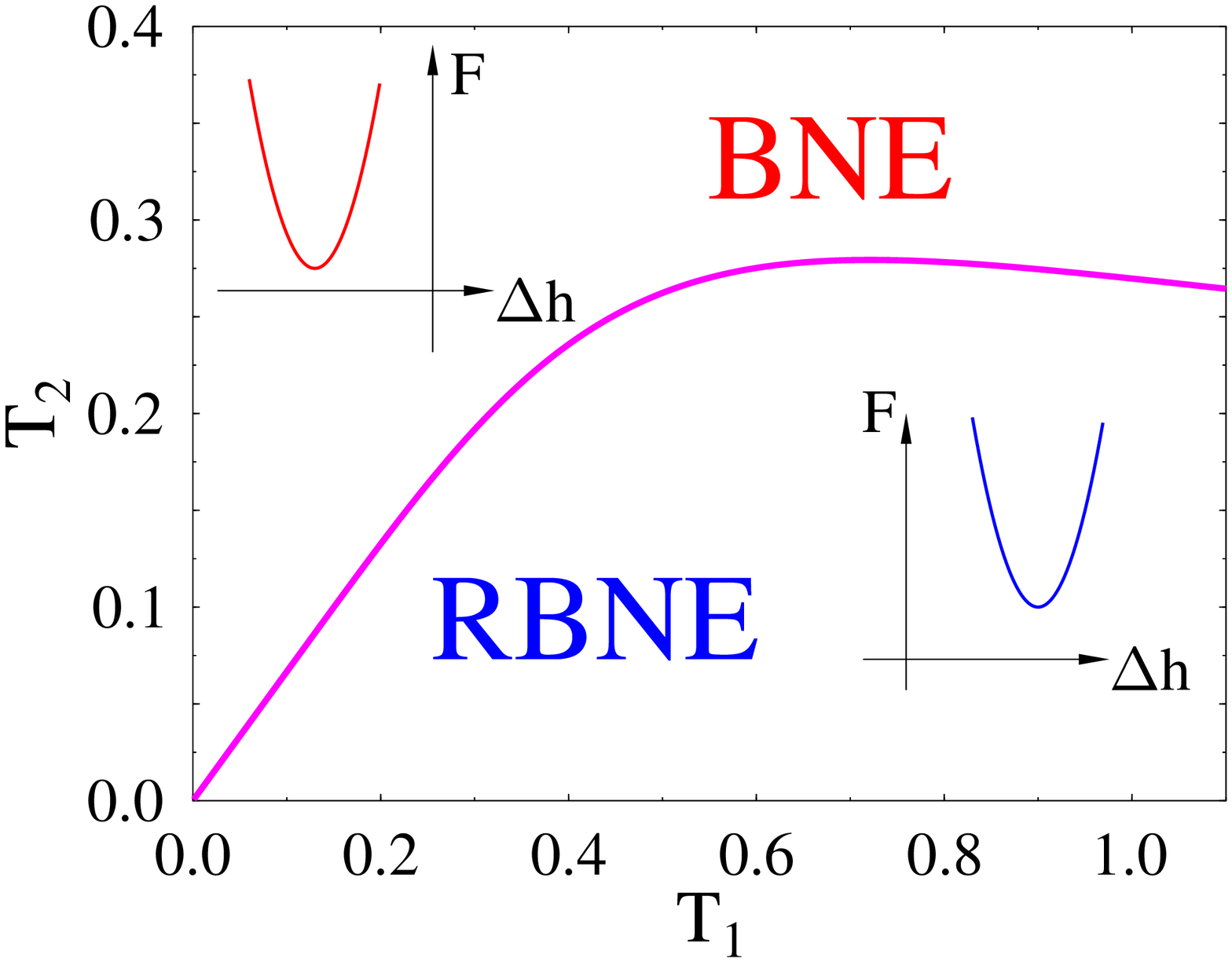,height=8.5cm,angle=-90}}
\vspace{-1.5cm}
\caption{
The analytically calculated diagram for segregation in the plane
$T_1=(m_1ga_0\beta_1)^{-1}$, $T_2=(m_2ga_0\beta_2)^{-1}$
in the three levels model described in the text
($\beta_i$ is the inverse configurational temperature of the grains
of type $i$).
The crossover from Brazil nut effect (BNE) to reverse Brazil nut effect (RBNE)
corresponds to the drift of the minimum of the free energy in the
inherent states space, $F$,
from negative to positive values of the vertical segregation parameter,
$\Delta h$.
}
\label{fig_diag_seg}
\end{figure}

We show now how the ensemble averages with the distribution (\ref{pr}) can be
analytically calculated for a simple toy model of a mixture of two
particles. The model can be visualized as a cell of the above hard sphere
pack, made of one large and two small grains seating on the vertices of a
square lattice, with three layers each of 2 sites. Interestingly, such a
three levels system is able to reproduce the general properties found
before. In particular, the resulting diagram for segregation is plotted in
Fig.\ref{fig_diag_seg} in the plane $T_1=(m_1ga_0\beta_1)^{-1}$,
$T_2=(m_2ga_0\beta_2)^{-1}$. It shows that for a given value of $T_1$ by
decreasing $T_2$ the system undergoes a crossover, at $T_2^*(T_1)$, from a
region where type-2 particles are on average above (BNE) to a region where
they are below (RBNE) type-1 particles. 
The structure of the diagram of Fig.\ref{fig_diag_seg}, 
for small values of $T_1$ and $T_2$, 
can be easily understood by looking at eq.(\ref{pr}): 
when the configurational temperature of small particles, $T_1$, is 
much smaller than the other, $T_2$, the 
potential energy of small grains is much smaller than that associated
to the large ones, therefore large grains segregate on top (BNE).
The opposite happens when $T_2$ is close to zero and much smaller than $T_1$. 
In particular, in this toy model the RBNE region cannot be met if 
$T_2$ is too large. 

The crossover observed in vertical segregation from BNE to RBNE
corresponds to the drift of the minimum of the free energy, $F=-\ln Z$,
from negative to positive values of $\Delta h$.
In this scenario, the pack undergoing a taps dynamics with
``amplitudes'' $T_{\Gamma}$ (and given ``duration'', $\tau_0$) at
stationarity, as for the data shown in Fig.\ref{fig_dh}, can be thus
thought to be following a given path, $T_2=T_2(T_1)$ (where
$T_1=T_1(T_{\Gamma},\tau_0)$), in the $(T_1,T_2)$ plane, crossing from the
BNE to the RBNE region.

In conclusion, we have shown that in the present
hard sphere binary mixture model Edwards' approach to describe
segregation holds if two thermodynamical parameters are introduced. 
In the case
considered here these two thermodynamical parameters play the role of
configurational temperatures related to the two types of particles. By
applying all the standard techniques of statistical mechanics of the
canonical distribution one can thus predict the complex segregation
patterns as found above. In particular we have shown on a hard sphere mixture 
on a lattice that the segregation phenomena in granular media 
can be studied applying standard statistical mechanics to a new 
Hamiltonian (\ref{effective}), which contains the original Hamiltonian of the 
grains plus a term which constrains the system to be in a inherent state.
  
In general for a more complex system one might expect more constraints to be 
imposed, leading to more than two thermodynamical parameters. 
In practice, the criteria to determine a priori the required parameters
can be not easily accessible. However more recently we have extended data for 
very low energy \cite{CFN} and found that only one thermodynamical parameter is necessary to describe the stationary state. This seems to be a general feature
\cite{fnc}. If this is the case, a Statistical Mechanics approach  with only one
thermodynamical variable may be feasible for low energy.  

\noindent 
This work was partially supported by the TMR-ERBFMRXCT980183,
INFM-PRA(HOP), MURST-PRIN 2000 and MURST-FIRB 2002. 
The allocation of computer resources from INFM
Progetto Calcolo Parallelo is acknowledged.

\vspace{-1.5cm}

\vspace{-2cm}

\begin{references}
\bibitem{rev_segr}  J.M. Ottino and D.V. Khakhar,
Ann. Rev. Fluid Mech. {\bf 32}, 55 (2000).
H.M. Jaeger, S.R. Nagel and R.P. Behringer,
Rev. Mod. Phys. {\bf 68}, 1259 (1996).
J. Bridgewater, 
Chem. Eng. Sci. {\bf 50}, 4081 (1995).

\bibitem{makse1}  H.A. Makse, S. Havlin, P.R. King, and H.E. Stanley,
Nature {\bf 386}, 379 (1997).
T. Shinbrot, A. Alexander, F.J. Muzzio,
Nature {\bf 397}, 675 (1999).
M.E. M\"{o}bius, B.E. Lauderdale, S.R. Nagel, H.M. Jaeger,
Nature {\bf 414}, 270 (2001).

\bibitem{rosato}  T. Rosato, F. Prinze, K.J. Standburg, R. Swendsen,
Phys. Rev. Lett. {\bf 58}, 1038 (1987).

\bibitem{bridgewater} J. Bridgewater, 
Powder Technol. {\bf 15}, 215 (1976).
J.C. Williams, 
Powder Technol. {\bf 15}, 245 (1976).

\bibitem{duran}  J. Duran, J. Rajchenbach, E. Clement,
Phys. Rev. Lett. {\bf 70}, 2431 (1993).

\bibitem{knight}  J. Knight, H. Jaeger, S. Nagel,
Phys. Rev. Lett. {\bf 70}, 3728 (1993).

\bibitem{luding}  D.C. Hong, P.V. Quinn, S. Luding,
Phys. Rev. Lett. {\bf 86}, 3423 (2001).
J.A. Both and D.C. Hong, 
Phys. Rev. Lett. {\bf 88}, 124301 (2002). 

\bibitem{shinbrot}  T. Shinbrot and F. Muzzio,
Phys. Rev. Lett. {\bf 81}, 4365 (1998).


\bibitem{Edwards1}  S.F. Edwards and R.B.S. Oakeshott, 
Physica A {\bf 157}, 1080 (1989).
A. Mehta and S.F. Edwards, 
Physica A {\bf 157}, 1091 (1989).
S.F. Edwards, in {\em Current Trends in the physics of
Materials}, (Italian Phys. Soc., North Holland, Amsterdam, 1990).

\bibitem{Dean}  A. Lef\`{e}vre, D. S. Dean,
Phys. Rev. Lett. {\bf 86}, 5639 (2001).

\bibitem{fnc} A. Coniglio and M. Nicodemi, 
Physica A {\bf 296}, 451 (2001).
A. Fierro, M. Nicodemi and A. Coniglio,
Europhys. Lett. 59 (5), 642 (2002).

\bibitem{brey}  J.J. Brey, A. Prados, B. S\'{a}nchez-Rey,
Physica A {\bf 275}, 310 (2000). 

\bibitem{kurchan}  
A. Barrat, J. Kurchan, V. Loreto and M. Sellitto,
Phys. Rev. Lett. {\bf 85}, 5034 (2000).
H.A. Makse and J. Kurchan,
Nature {\bf 415}, 614 (2002).

\bibitem{berg}  J. Berg, S. Franz and M. Sellitto,
Eur. Phys. J. B {\bf 26}, 349 (2002).                                           

\bibitem{lefevre}
The possibility of introducing more than one
thermodynamical parameter has been also suggested in \cite{berg} and
very recently discussed in the context of a
Constrained Ising Chain by A. Lef\`{e}vre,
cond-mat/0202376.
  
\bibitem{NCH} M. Nicodemi, A. Coniglio, H.J. Herrmann,
Phys. Rev. E {\bf 55}, 3962 (1997).
A. Coniglio, M. Nicodemi, 
J. Phys.: Cond. Matt. {\bf 12}, 6601 (2000).
M. Nicodemi, Physica A {\bf 285},  267 (2000).

\bibitem{Stillinger}
F.H. Stillinger T.A. Weber, 
Phys. Rev. A {\bf 25}, 978 (1982).
S. Sastry, P.G. Debenedetti, F.H. Stillinger,
Nature {\bf 393}, 554 (1998).
B. Coluzzi, G. Parisi and P. Verrocchio, Phys. Rev. Lett. {\bf 84}, 306 (2000).
F. Sciortino, W. Kob, P. Tartaglia, Phys. Rev. Lett. {\bf 83}, 3214 (1999).
F. Sciortino and P. Tartaglia, Phys. Rev. Lett. {\bf 86}, 107 (2001). 


\bibitem{nota_num_p}  We considered two kinds of systems: a box of
size $L=20$, with $N_{1}=120$ small and $N_{2}=40$ large grains; and
a box of size $L=10$, with $N_{1}=50$ and $N_{2}=25$. Our data are 
averaged on up to 16384 realizations. We also considered sizes $L$ up to 32, 
$N_1$ up to 480 and $N_2$ up to 160. 

\bibitem{mimmo} M. Nicodemi, Phys. Rev. Lett. {\bf 82}, 3734 (1999). 

\bibitem{CFN} A. Coniglio, A. Fierro, M. Nicodemi, in preparation.

\end{references}
\end{document}